\documentclass{article}
\usepackage[final]{aaj2026}
\usepackage[numbers]{natbib}
\usepackage{graphicx}
\usepackage[utf8]{inputenc}
\usepackage[T1]{fontenc}
\usepackage{hyperref}
\usepackage{url}
\usepackage{booktabs}
\usepackage{amsfonts,amsmath,amssymb}
\usepackage{nicefrac}
\usepackage{microtype}
\usepackage{xcolor}
\usepackage{enumitem}
\usepackage{tcolorbox}
\usepackage{multirow}
\usepackage{forest}

\definecolor{navy}{HTML}{1E293B}
\definecolor{lgray}{HTML}{F1F5F9}

\newcommand{\dcos}{d_{\mathrm{cos}}}
\newcommand{\geogap}{\textsc{GeoGap}}
\newcommand{\geogapg}{\textsc{GeoGap-G}}
\newcommand{\geogapgt}{\textsc{GeoGap-GT}}

\title{Detecting Underspecification in Software Requirements\\via $k$-NN Coverage Geometry\thanks{%
  \href{https://uenian33.github.io/gaplens/}{Project Page}\quad$\cdot$\quad
  \href{https://github.com/GPT-Laboratory/DiveAI_Case-16-Visualizing-the-Requirements-Space/tree/data_analysis_research_questions}{Code}\quad$\cdot$\quad
  \href{https://www.overleaf.com/read/szffnrcbxcpg\#ecc26d}{Overleaf}%
}}

\author{%
  Wenyan Yang \\
  Tampere University \\
  \texttt{wenyan.yang@tuni.fi} \\
  \And
  Tom\'{a}\v{s} Janovec \\
  Tampere University \\
  \texttt{tomas.janovec@tuni.fi} \\
  \And
  Samantha Bavautdin \\
  Tampere University \\
  \texttt{samantha.bavautdin@tuni.fi} \\
}

\begin{document}
\maketitle

\begin{abstract}
We propose \geogap{}, a geometric method for detecting missing requirement types in software specifications. The method represents each requirement as a unit vector via a pretrained sentence encoder, then measures coverage deficits through $k$-nearest-neighbour distances z-scored against per-project baselines. Three complementary scoring components---per-point geometric coverage, type-restricted distributional coverage, and annotation-free population counting---fuse into a unified gap score controlled by two hyperparameters. On the PROMISE NFR benchmark, \geogap{} achieves 0.935 AUROC for detecting completely absent requirement types in projects with $N \geq 50$ requirements, matching a ground-truth count oracle that requires human annotation. Six baselines confirm that each pipeline component---per-project normalisation, neural embeddings, and geometric scoring---contributes measurable value.
\end{abstract}

\section{Introduction}\label{sec:intro}

Software requirements specifications are routinely incomplete. A healthcare system may specify authentication and audit logging but omit backup and disaster-recovery requirements that every peer system addresses. Detecting such \emph{underspecification} before design commits resources prevents costly rework.

No universal checklist of required concerns exists: what a project \emph{should} address depends on its domain, scale, and quality priorities. We therefore frame the problem as a \textbf{coverage comparison}: given a corpus of peer projects with known requirement types, identify regions of concern the corpus covers but the target project does not.

We propose \geogap{}, which represents each requirement as a unit vector on the hypersphere $\mathbb{S}^{d-1}$ via a pretrained sentence encoder, then measures coverage deficits via $k$-nearest-neighbour distances. The method has three scoring components:

\begin{itemize}[nosep]
  \item \textbf{Geometric coverage} ($\Psi_{\mathrm{geo}}$): for each corpus point, is the target's nearest requirement unusually far?
  \item \textbf{Type-restricted coverage} ($\Psi_{\mathrm{type}}$): the same question, restricted to same-type requirements only.
  \item \textbf{Population count} ($\Psi_{\mathrm{pop}}$): does the target have fewer requirements of each type than peer projects?
\end{itemize}

\begin{figure}[ht]
\centering
\includegraphics[width=\textwidth]{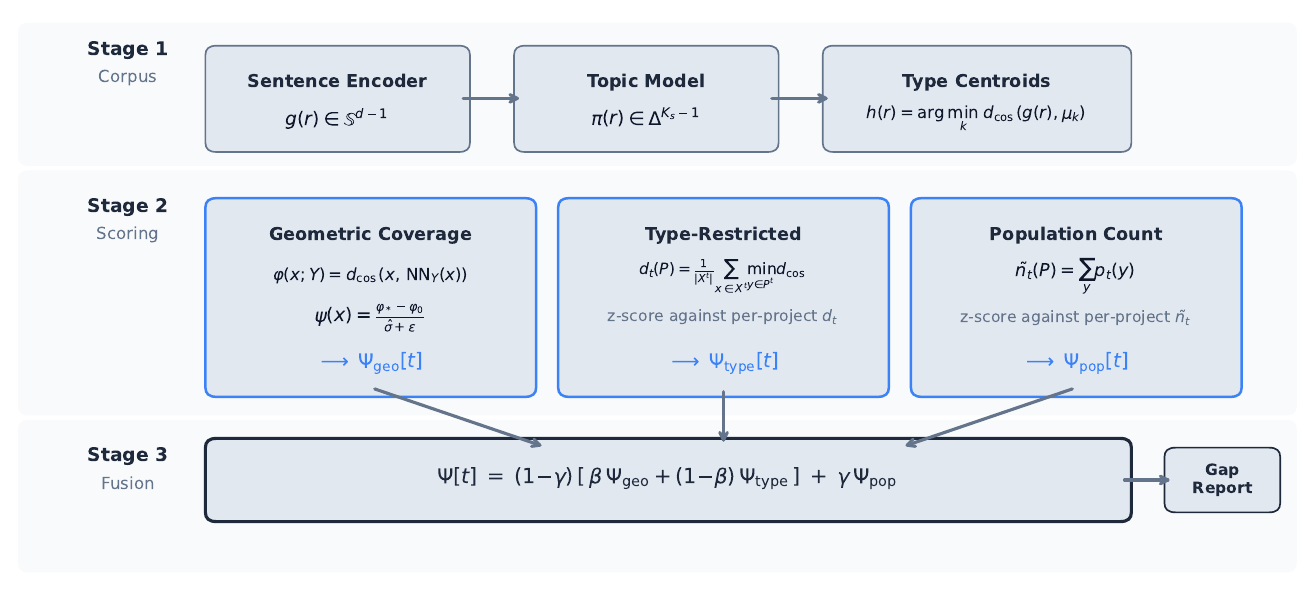}
\caption{The \geogap{} pipeline. Stage~1 builds the reference corpus (embeddings, topics, type centroids). Stage~2 computes three complementary gap scores. Stage~3 fuses them and produces a structured gap report with heatmaps.}
\label{fig:pipeline}
\end{figure}

\noindent These fuse into a single score via two hyperparameters, then aggregate onto a human-readable (category$\times$topic) grid. Figure~\ref{fig:pipeline} shows the pipeline.

\section{Problem Definition}\label{sec:problem}

\subsection{Dataset}

We use the PROMISE NFR dataset~\citep{clelandhuang2006}: $N = 621$ natural-language requirement sentences from $M = 15$ software projects. Each requirement $r$ has a project identifier $\mathrm{pid}(r) \in \{P_1, \dots, P_{15}\}$ and a quality-attribute type from $K_t = 12$ classes (Availability, Fault Tolerance, Functional, Legal, Look~\&~Feel, Maintainability, Operability, Performance, Portability, Scalability, Security, Usability). Project sizes range from 12 to 92; six have $N_j \geq 50$.

\subsection{Gap Definition}

Let $P_*$ be a target project and $\mathcal{D} = \{P_1, \dots, P_M\}$ a reference corpus. A sentence encoder $g$ maps each requirement to a unit vector $g(r) \in \mathbb{S}^{d-1}$ ($d = 1024$).

\begin{tcolorbox}[colback=lgray, colframe=navy!60, left=6pt, right=6pt, top=4pt, bottom=4pt]
\textbf{Gap.} A region of $\mathbb{S}^{d-1}$ where the corpus has substantial coverage (many points from multiple projects) but the target has little or no coverage.
\end{tcolorbox}

\noindent We address three questions:
\begin{description}[nosep, leftmargin=1.5em]
  \item[RQ1] Which types does the entire corpus cover but $P_*$ does not? (Structural gaps)
  \item[RQ2] Which types do similar projects cover but $P_*$ does not? (Domain-specific gaps)
  \item[RQ3] Which requirements in $P_*$ have no analogue in the corpus? (Novelties)
\end{description}

\section{Representation}\label{sec:repr}

\subsection{Sentence Embedding}

Every requirement is encoded using Qwen3-Embedding-0.6B~\citep{qwen3embedding2025} ($d = 1024$) in plain passage mode: $g(r) \in \mathbb{S}^{d-1}$, $\|g(r)\| = 1$. Cosine distance: $\dcos(u,v) = 1 - u^\top v$.

\subsection{Type Centroids}

For each type $k$, a centroid is computed from PROMISE ground-truth labels:
\begin{equation}\label{eq:centroid}
  \boldsymbol{\mu}_k = \frac{\sum_{r \in \mathcal{R}_k} g(r)}{\bigl\|\sum_{r \in \mathcal{R}_k} g(r)\bigr\|}, \qquad h(r) = \arg\min_{k} \dcos\!\bigl(g(r), \boldsymbol{\mu}_k\bigr).
\end{equation}
Hard assignment $h(r)$ achieves 78.3\% accuracy.

\subsection{Soft Topic Distributions}

BERTopic~\citep{grootendorst2022bertopic} discovers latent topics via UMAP~\citep{mcinnes2018umap} $\to$ HDBSCAN~\citep{mcinnes2017hdbscan} $\to$ c-TF-IDF. Each requirement receives a soft distribution $\boldsymbol{\pi}(r) \in \Delta^{K_s-1}$, ensuring every requirement contributes to every topic (no outlier loss).

\section{Method}\label{sec:method}

\begin{figure}[ht]
\centering
\includegraphics[width=\textwidth]{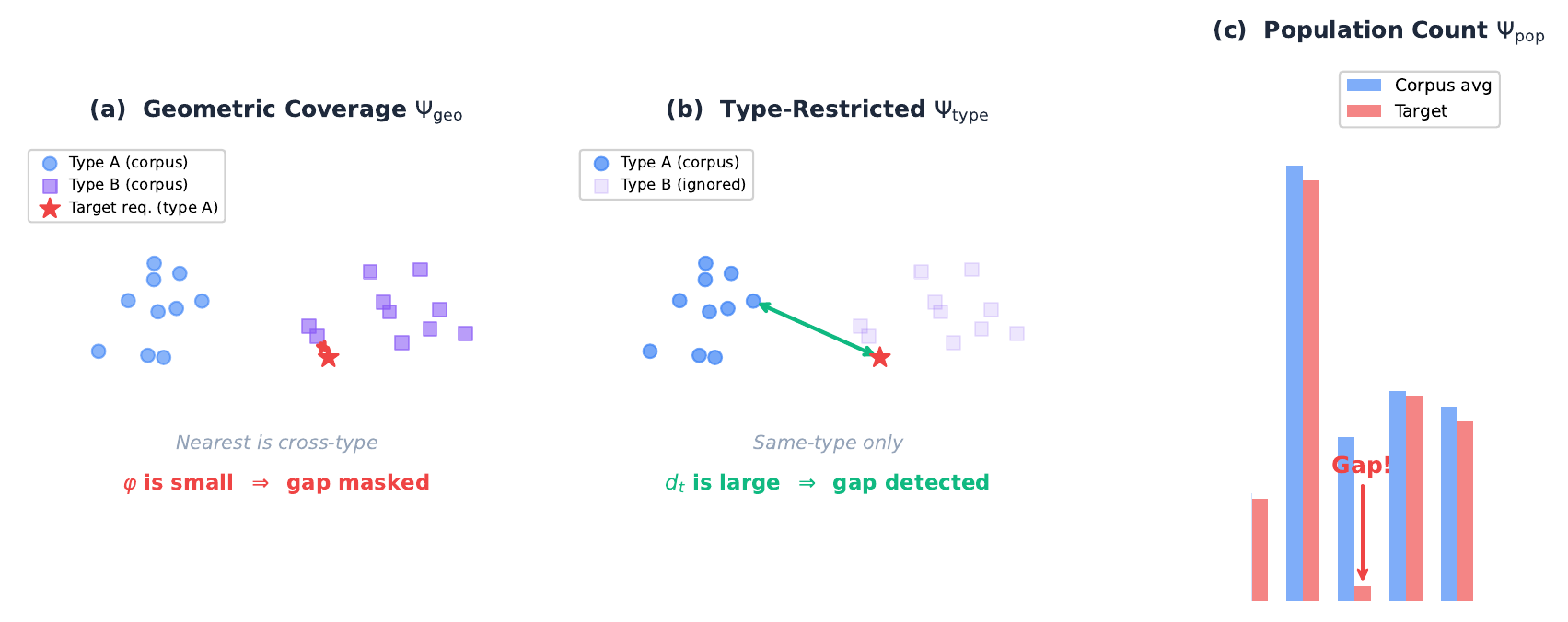}
\caption{Three scoring components. \textbf{(a)}~Geometric coverage ($\Psi_{\mathrm{geo}}$): unrestricted $k$-NN finds a cross-type neighbour, masking the gap. \textbf{(b)}~Type-restricted coverage ($\Psi_{\mathrm{type}}$): restricting to same-type neighbours reveals the gap. \textbf{(c)}~Population count ($\Psi_{\mathrm{pop}}$): the target has far fewer Performance requirements than the corpus average.}
\label{fig:three-layers}
\end{figure}

\geogap{} computes a gap score $\Psi[t]$ for each requirement type $t$ by fusing three complementary signals. Figure~\ref{fig:three-layers} illustrates what each captures.

\subsection{Geometric Coverage ($\Psi_{\mathrm{geo}}$)}\label{sec:geo}

\begin{figure}[ht]
\centering
\includegraphics[width=0.85\textwidth]{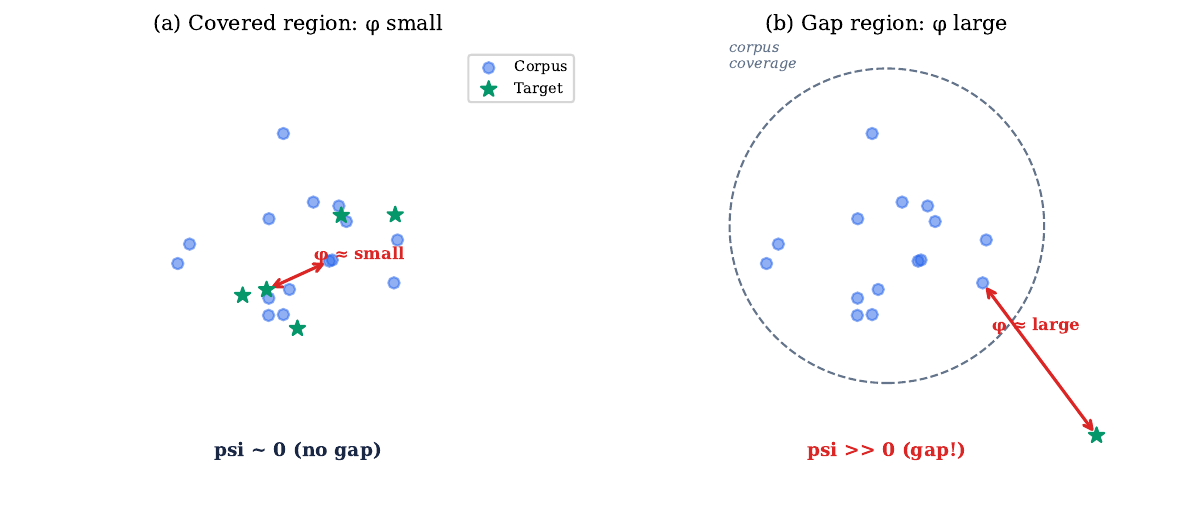}
\caption{Geometric intuition. (a)~A covered region: the target has requirements near corpus point $x$, so $\varphi$ is small. (b)~A gap region: the target's nearest requirement is far from $x$, so $\varphi$ is large. Per-project normalisation (Figure~\ref{fig:normalisation}) determines whether a large $\varphi$ is unusual or merely reflects inherent sparsity.}
\label{fig:intuition}
\end{figure}

For each corpus point $x$, the coverage distance measures how far the target's nearest requirement is:
\begin{equation}\label{eq:phi}
  \varphi(x;\, \mathbf{Y}) = \frac{1}{k}\sum_{i=1}^{k} \dcos\!\bigl(x, y_{(i)}\bigr),
\end{equation}
where $y_{(1)}, \dots, y_{(k)}$ are $x$'s $k$ nearest neighbours in the target. A large $\varphi$ suggests a coverage deficit---but some regions are inherently sparse. Thresholding raw $\varphi$ would flag the same regions for every target. Figure~\ref{fig:intuition} illustrates the distinction.

\begin{figure}[ht]
\centering
\includegraphics[width=\textwidth]{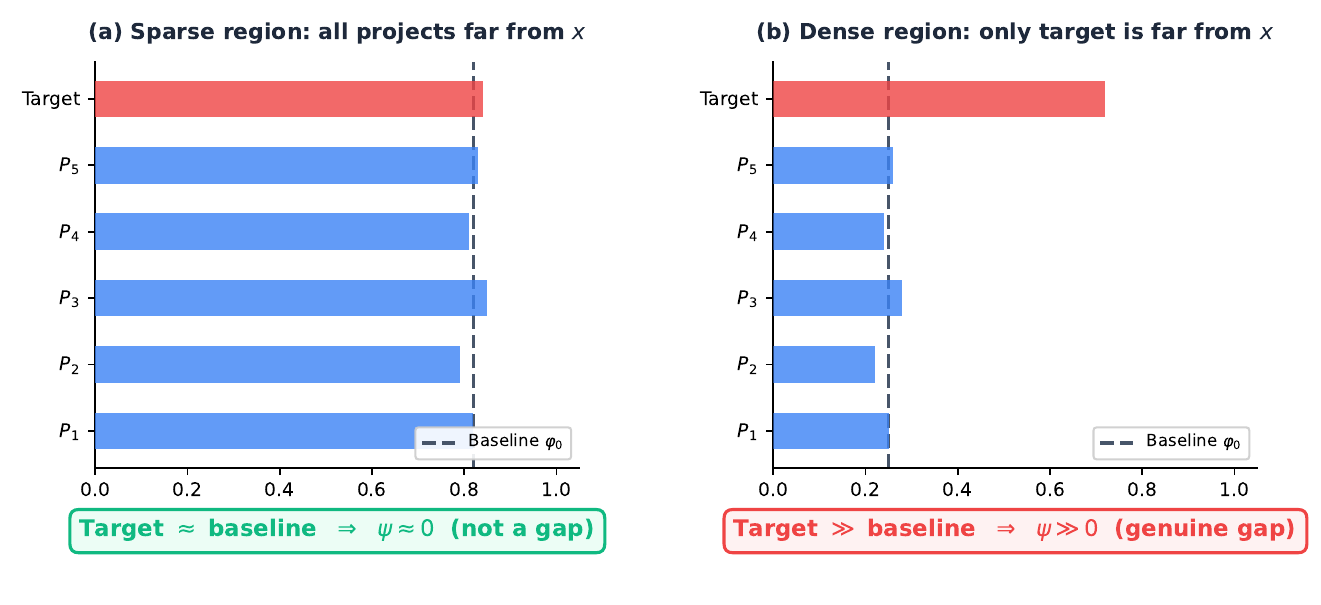}
\caption{Per-project normalisation. (a)~In a sparse region, all projects are equally far from $x$, so the target is typical ($\psi \approx 0$, not a gap). (b)~In a dense region, training projects are close but the target is far ($\psi \gg 0$: genuine gap).}
\label{fig:normalisation}
\end{figure}

\paragraph{Per-project baseline normalisation.} The core idea: instead of asking ``is $\varphi$ large?'', ask ``is it unusually large compared to what a typical individual project achieves?'' (Figure~\ref{fig:normalisation}).

Compute $\varphi(x; \mathbf{Y}_j)$ for each training project individually:
\begin{equation}\label{eq:phi0}
  \varphi_0(x) = \frac{1}{M}\sum_{j=1}^{M} \varphi(x; \mathbf{Y}_j), \qquad \hat{\sigma}(x) = \mathrm{std}_j\{\varphi(x; \mathbf{Y}_j)\}.
\end{equation}
The normalised gap score is:
\begin{equation}\label{eq:psi}
  \boxed{\;\psi(x) = \operatorname{clip}\!\left(\frac{\varphi(x; \mathbf{Y}_*) - \varphi_0(x)}{\hat{\sigma}(x) + \varepsilon},\; -5,\; 5\right)\;}
\end{equation}

Per-point scores are marginalised to type level via cell aggregation (Section~\ref{sec:aggregation}):
$\Psi_{\mathrm{geo}}[t] = \sum_s W[t,s] \cdot \Psi_{\mathrm{cell}}[t,s] \;/\; \sum_s W[t,s]$.

\paragraph{Why $k = 1$.} On $\mathbb{S}^{d-1}$ with $d = 1024$, concentration of measure~\citep{ledoux2001concentration} causes all but the nearest neighbour to lie at approximately the same ambient distance. The $k$-sensitivity experiment (Section~\ref{sec:ksens}) confirms: \geogapg{} AUROC decays from 0.87 ($k{=}1$) to 0.39 ($k{=}20$).

\subsection{Type-Restricted Coverage ($\Psi_{\mathrm{type}}$)}\label{sec:typerestr}

The geometric score uses unrestricted $k$-NN: the target's closest requirement to $x$ may belong to a different type, masking a type-level gap. The type-restricted score restricts the search to same-type requirements:
\begin{equation}\label{eq:dt}
  d_t(P) = \frac{1}{|\mathbf{X}^t|} \sum_{x \in \mathbf{X}^t} \min_{y \in P^t} \dcos(x, y),
\end{equation}
z-scored against per-project baselines:
\begin{equation}\label{eq:psi-type}
  \boxed{\;\Psi_{\mathrm{type}}[t] = \operatorname{clip}\!\left(\frac{d_t(P_*) - \bar{d}_t}{\hat{\sigma}_{d_t} + \varepsilon},\; -5,\; 5\right)\;}
\end{equation}

Key properties: (i)~amplifies partial-removal signal by ${\sim}1.9\times$; (ii)~insensitive to $k$ because $d_t$ averages over all reference points of type $t$.

\subsection{Population Count ($\Psi_{\mathrm{pop}}$)}\label{sec:pop}

Our baseline experiments (Section~\ref{sec:baselines}) revealed that counting requirements per type with ground-truth labels achieves 0.933 AUROC---population statistics carry signal that geometric coverage misses. We capture this without annotation via calibrated Gibbs soft assignment:
\begin{equation}\label{eq:soft-count}
  \tilde{n}_t(P) = \sum_{y \in P} p_t(y), \quad p_t(y) = \frac{\exp(-\dcos(y, \boldsymbol{\mu}_t)/T)}{\sum_{t'}\exp(-\dcos(y, \boldsymbol{\mu}_{t'})/T)},
\end{equation}
with $T^* = 0.021$ calibrated so $\mathbb{E}[\max_t p_t(y)]$ matches the hard-assignment accuracy (0.783). The population score is:
\begin{equation}\label{eq:psi-pop}
  \boxed{\;\Psi_{\mathrm{pop}}[t] = \operatorname{clip}\!\left(\frac{\bar{\tilde{n}}_t - \tilde{n}_t(P_*)}{\hat{\sigma}_{\tilde{n}_t} + \varepsilon},\; -5,\; 5\right)\;}
\end{equation}

The count signal is linear in the removal fraction ($5.3\times$ more dynamic range than the geometric score).

\subsection{Fusion}\label{sec:fusion}

\begin{equation}\label{eq:fusion}
  \boxed{\;\Psi[t] = (1 - \gamma)\bigl[\beta\,\Psi_{\mathrm{geo}}[t] + (1-\beta)\,\Psi_{\mathrm{type}}[t]\bigr] + \gamma\,\Psi_{\mathrm{pop}}[t]\;}
\end{equation}

\begin{figure}[ht]
\centering
\includegraphics[width=0.85\textwidth]{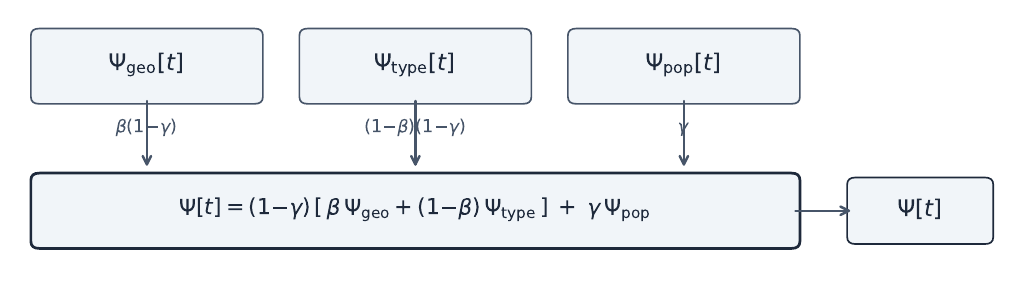}
\caption{Score fusion via two hyperparameters. Three configurations are studied (Table~\ref{tab:configs}).}
\label{fig:fusion}
\end{figure}

\begin{table}[ht]
\centering\small
\caption{Algorithm configurations. \geogap{} (full) is recommended.}
\label{tab:configs}
\begin{tabular}{@{}llccl@{}}
\toprule
\textbf{Name} & \textbf{Active scores} & $\beta$ & $\gamma$ & \textbf{Description} \\
\midrule
\geogapg{} & $\Psi_{\mathrm{geo}}$ only & 1.0 & 0.0 & Geometric only (ablation) \\
\geogapgt{} & $\Psi_{\mathrm{geo}} + \Psi_{\mathrm{type}}$ & 0.7 & 0.0 & Adds type-restricted ($k$-robust) \\
\geogap{} & All three & 0.7 & 0.1 & Full method \\
\bottomrule
\end{tabular}
\end{table}

\subsection{Cell Aggregation}\label{sec:aggregation}

Per-point scores aggregate onto a (type$\times$topic) grid:
\begin{equation}\label{eq:Psi-cell}
  \Psi_{\mathrm{cell}}[t,s] = \frac{\sum_{x: h(x)=t} \pi_s(x) \cdot \psi(x)}{\sum_{x: h(x)=t} \pi_s(x)}.
\end{equation}
Occupancy: $\widetilde{M}_*[t,s] = \sum_{y \in P_*: h(y)=t} \pi_s(y)$, with $\sum_{t,s} \widetilde{M}_* = N_*$.

\begin{figure}[ht]
\centering
\includegraphics[width=\textwidth]{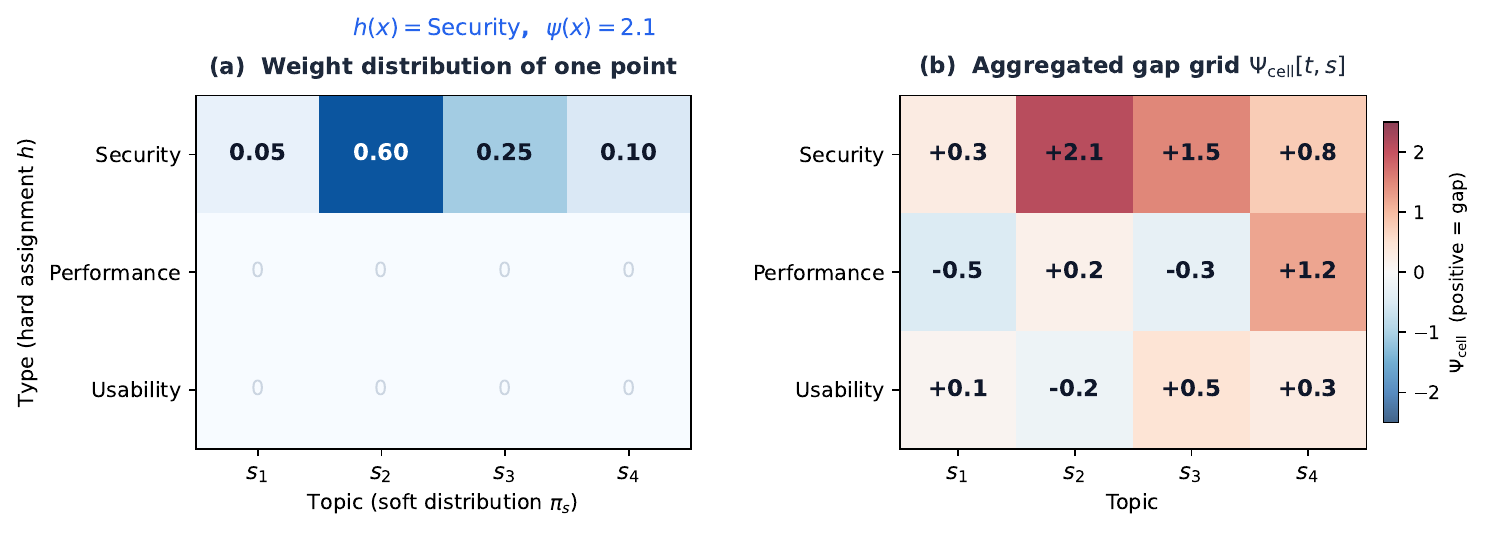}
\caption{Soft cell aggregation. (a)~A reference point distributes its score across topic columns via $\boldsymbol{\pi}$. (b)~Aggregated grid: red = gap, blue = surplus.}
\label{fig:cell-agg}
\end{figure}

\subsection{Dual-Mode Comparison}\label{sec:dual}

The same per-project distances support two modes: \textbf{Mode~A} (uniform weights, RQ1) and \textbf{Mode~B} (similarity-weighted with $w_j \propto \exp(-\dcos(\boldsymbol{\mu}_*, \boldsymbol{\mu}_j)/\tau)$, optionally using optimal transport~\citep{cuturi2013sinkhorn}, RQ2).

\section{Experiments}\label{sec:eval}

We lack ground-truth gap labels, so all experiments use \textbf{synthetic gap injection}: remove requirements from well-covered types, then check whether the detector ranks depleted types as top gaps. All experiments use LOO-CV across 15 projects.

\begin{figure}[ht]
\centering
\includegraphics[width=\textwidth]{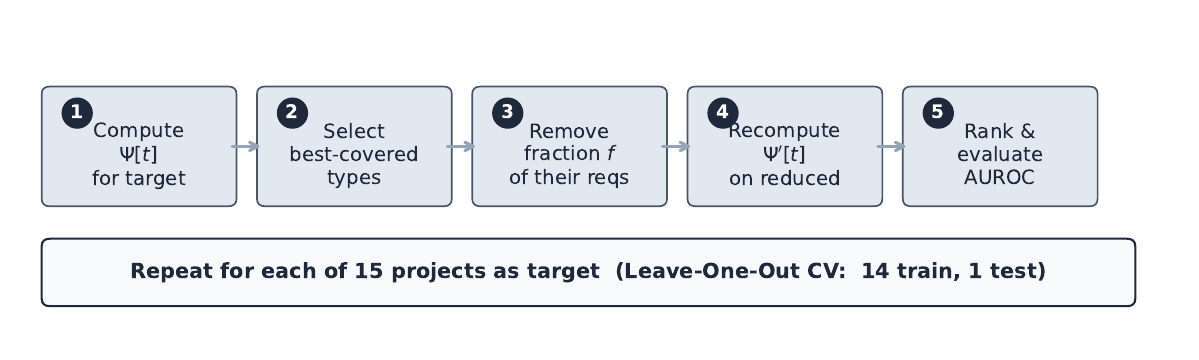}
\caption{Evaluation protocol. Each of 15 projects takes a turn as target; well-covered types are identified, their requirements removed, and the detector is evaluated on whether it ranks those types as gaps.}
\label{fig:loocv}
\end{figure}

\subsection{Type-Level Gap Detection}\label{sec:exp1}

\textbf{Question.} Can the detector identify which requirement types are completely absent from a project?

\textbf{Protocol.} For each LOO fold: compute $\Psi[t]$ for all 12 types; select 1--3 best-covered types; remove all their requirements ($f = 1.0$); recompute $\Psi'[t]$; rank types by $\Psi'$ descending; evaluate whether the depleted types rank near the top.

\textbf{Metrics.}
\emph{AUROC} (Area Under the ROC Curve) measures ranking quality: can the detector rank truly-depleted types above non-depleted ones?  AUROC $= 1.0$ means perfect separation; $0.5$ means random.  This is the primary metric because gap detection is fundamentally a \emph{ranking} task---the practitioner examines top-ranked types first.
\emph{MRR} (Mean Reciprocal Rank) measures how early the first true gap appears: $\mathrm{MRR} = \frac{1}{n}\sum_{i=1}^n \frac{1}{r_i}$, where $r_i$ is the rank of the $i$-th true gap.  A high MRR means the practitioner finds a real gap quickly.

\begin{table}[ht]
\centering\small
\caption{Type-level detection ($f{=}1.0$, $k{=}1$). \geogap{} achieves 0.935 AUROC for $N \geq 50$.}
\label{tab:type-main}
\begin{tabular}{@{}lccc@{}}
\toprule
& \textbf{AUROC} ($N{\geq}50$) & \textbf{AUROC} (All) & \textbf{MRR} (All) \\
\midrule
\geogapg{} (geometric only) & $0.871 \pm 0.22$ & $0.764 \pm 0.26$ & $0.453$ \\
\geogapgt{} (+ type-restricted) & $0.902 \pm 0.15$ & $0.795 \pm 0.21$ & $0.422$ \\
\geogap{} (full) & $\mathbf{0.935 \pm 0.14}$ & $\mathbf{0.801 \pm 0.20}$ & $\mathbf{0.472}$ \\
\bottomrule
\end{tabular}
\end{table}

\textbf{Finding.} The full \geogap{} improves over geometric-only by $+0.064$ AUROC for large projects ($N \geq 50$), with gains concentrated on \geogapg{}'s weakest cases (PROMISE\_8: $+0.148$; PROMISE\_4: $+0.200$).  The MRR of 0.472 means that, on average, the first correctly identified gap appears around rank~2.

\subsection{Fraction Monotonicity}\label{sec:exp2}

\textbf{Question.} Does the detector's output scale with the severity of the gap?  A reliable detector should produce higher scores when \emph{more} requirements are removed.

\textbf{Protocol.} Same as above, with removal fractions $f \in \{1.0, 0.75, 0.50\}$ (complete, three-quarter, and half removal).

\textbf{Metric.} AUROC at each fraction.  We expect monotonicity: $\mathrm{AUROC}(f{=}1.0) > \mathrm{AUROC}(f{=}0.75) > \mathrm{AUROC}(f{=}0.50)$.  A violation would indicate the detector responds to something other than gap magnitude.

\begin{figure}[ht]
\centering
\includegraphics[width=0.5\textwidth]{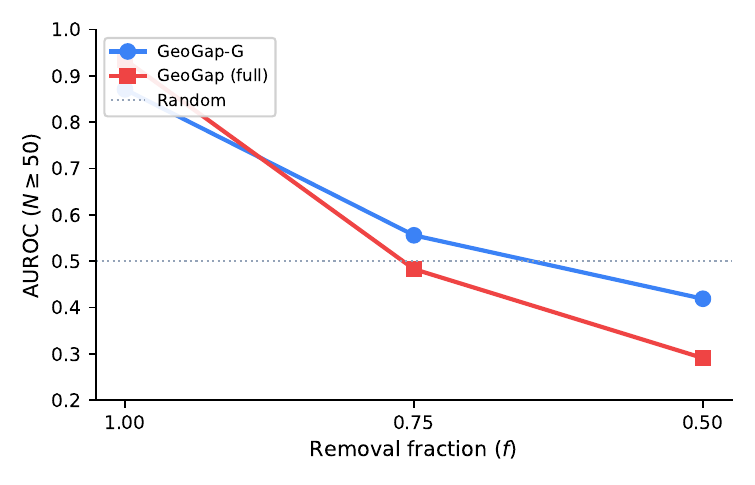}
\caption{AUROC vs.\ removal fraction ($N \geq 50$). Monotonicity holds; the steep drop from $f{=}1.0$ to $f{=}0.75$ confirms the method primarily detects complete absences.}
\label{fig:mono}
\end{figure}

\textbf{Finding.} Monotonicity is satisfied for both configurations, confirming the detector's output tracks gap severity.  However, the steep drop ($0.935 \to 0.483$ for \geogap{}) reveals a structural limitation: in $d{=}1024$ with small point clouds, removing 25\% of a type's requirements barely shifts the $k$-NN geometry---surviving requirements still provide coverage for most reference points.  The method is reliable for detecting \emph{complete} absences but not partial underspecification.

\subsection{$k$-Sensitivity}\label{sec:ksens}

\textbf{Question.} How sensitive is detection to the number of nearest neighbours $k$?  Is the gap signal strictly local (carried by the 1-NN only), and does \geogap{} eliminate this sensitivity?

\textbf{Protocol.} Type-level injection with $f = 1.0$, varying $k \in \{1, 3, 5, 10, 20\}$.

\textbf{Metric.} AUROC at each $k$.  A monotonic decay with $k$ proves the signal is local (concentration of measure on $\mathbb{S}^{d-1}$).  A flat curve proves the method is $k$-robust.  A random detector would show a constant line at ${\sim}0.5$ regardless of $k$---any deviation from flatness is evidence of a real signal.

\begin{figure}[ht]
\centering
\includegraphics[width=0.55\textwidth]{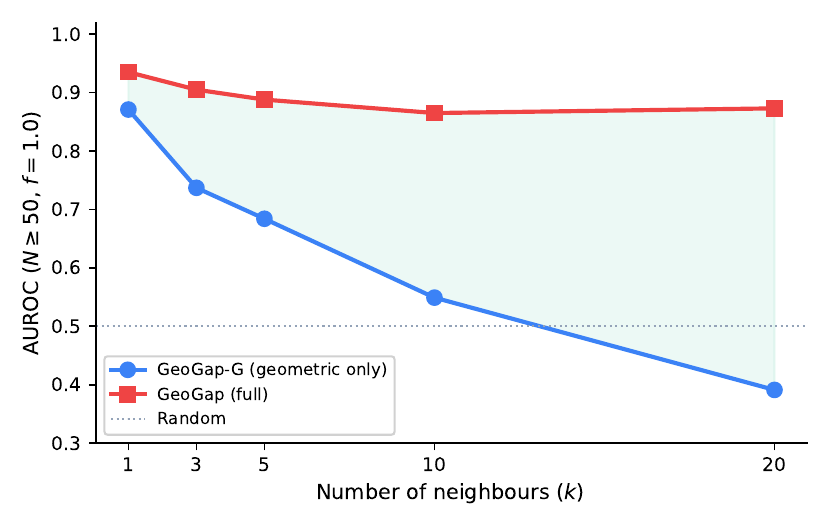}
\caption{$k$-sensitivity ($N \geq 50$, $f{=}1.0$). \geogapg{} decays from 0.87 to 0.39 (strictly local signal). \geogap{} stays above 0.87 at all $k$ (robust). A random detector would be flat at 0.5.}
\label{fig:ksens}
\end{figure}

\textbf{Finding.} \geogapg{} decays monotonically ($0.871 \to 0.391$), confirming the gap signal is strictly local: only the 1-nearest neighbour carries coverage information on $\mathbb{S}^{1023}$.  This monotonic decay is also the \emph{strongest evidence that the detector captures a real signal}---a noise-based detector cannot produce this pattern.

The full \geogap{} is nearly flat ($0.935 \to 0.873$, a drop of only 0.062 across a $20\times$ change in $k$).  This is the most important structural improvement: \geogapg{} has a critical hyperparameter ($k{=}1$) that must be set exactly; \geogap{} does not.  The robustness arises because $\Psi_{\mathrm{type}}$ and $\Psi_{\mathrm{pop}}$ aggregate across all reference points per type \emph{before} computing the z-score.

\subsection{Permutation Test}\label{sec:perm}

\textbf{Question.} Are the observed per-project AUROCs statistically significant, or could they arise by chance given the small number of types ($K_t = 12$)?

\textbf{Protocol.} For each project, run 1000 random permutations: label 3 types as ``targets'' without removing anything, compute AUROC of the actual $\Psi[t]$ scores against these random labels. This builds a null distribution of AUROC under $H_0$: ``scores are independent of which types are targets.''

\textbf{Metric.} The null 95th percentile (${\approx}0.83$ for $K_t{=}12$ with 3 targets).  A project is \emph{statistically significant} if its observed AUROC from the injection experiment exceeds this threshold.  This tells us whether the detection result for that project could be explained by chance.

\textbf{Finding.} Under \geogap{}, 9 of 15 projects exceed the null 95th percentile: all 6 large projects ($N \geq 50$) and 3 small ones.  The 6 non-significant projects all have $N < 50$, confirming that detection power depends on project size---small point clouds on $\mathbb{S}^{1023}$ provide insufficient geometric contrast.

\subsection{Cell-Level Localisation}\label{sec:exp5}

\textbf{Question.} Can the detector pinpoint gaps to specific (type~$\times$~topic) cells, not just types?  This tests whether the method can say ``you are missing Security requirements \emph{about encryption}'' rather than just ``you are missing Security requirements.''

\textbf{Protocol.} Same injection protocol, but operating on the full cell grid ($K_t \times K_s \approx 96$ cells) with 5 target cells.

\textbf{Metric.} AUROC over all cells.  The search space is ${\sim}8\times$ larger than type-level, requiring finer geometric discrimination.

\textbf{Finding.} AUROC = 0.51 (indistinguishable from random). With 621 corpus points spread across ${\sim}96$ cells (${\sim}6.5$ per cell on average), removing requirements from a single cell barely shifts the $k$-NN geometry.  This is an important negative result: it bounds the method's spatial resolution at the current corpus size.  A corpus 5--10$\times$ larger may enable cell-level detection.

\subsection{Whole-Project Holdout}\label{sec:exp6}

\textbf{Question.} If an entire project is removed from the training corpus, do its dominant types now appear as gaps for similar target projects?  This tests corpus-level sensitivity rather than target-level sensitivity.

\textbf{Protocol.} Remove $P_j$ from training; for each other project $P_k$ as target, check whether the types where $P_j$ contributed $>$20\% of corpus mass rank as gaps.

\textbf{Metric.} AUROC over the 12 types, treating dominant types of the removed project as positives.

\textbf{Finding.} 84 valid project pairs; mean AUROC = 0.50 (random). With $M{=}15$ projects, removing one shifts the per-project baseline $\varphi_0(x)$ by only ${\sim}7\%$---well within the noise.  This protocol would require a larger, more diverse corpus where individual projects have distinctive type signatures.

\subsection{Baseline Comparison}\label{sec:baselines}

\textbf{Question.} Does each component of the pipeline contribute measurable value, or could a simpler method achieve comparable results?  Without this experiment, a reviewer could argue the entire pipeline is unnecessary.

\textbf{Protocol.} Six baselines are evaluated on the \emph{identical} injection protocol ($f{=}1.0$, LOO-CV) as our method.  Each ablates exactly one pipeline component (Table~\ref{tab:baselines}).

\textbf{Metric.} AUROC, compared head-to-head.  A baseline that matches or exceeds \geogap{} would demonstrate the ablated component is unnecessary.

\begin{table}[ht]
\centering\small
\caption{Baselines ($f{=}1.0$, $k{=}1$). Each ablates one pipeline component.}
\label{tab:baselines}
\begin{tabular}{@{}llcc@{}}
\toprule
\textbf{Method} & \textbf{What it tests} & \textbf{AUROC} ($N{\geq}50$) & \textbf{AUROC} (All) \\
\midrule
Random             & Null                             & 0.345 & 0.309 \\
Count (GT labels)  & Is ML needed? (unfair oracle)    & \underline{0.933} & 0.795 \\
TF-IDF + $k$-NN   & Are neural embeddings needed?     & 0.615 & 0.549 \\
Classifier~\citep{hey2020norbert} & Is geometric scoring needed? & 0.673 & 0.573 \\
MMD/type~\citep{gretton2012kernel} & Is normalisation needed? & 0.265 & 0.230 \\
Centroid dist.     & Is normalisation needed?          & 0.284 & 0.266 \\
\midrule
\geogapg{} (ours)   & Geometric only                  & 0.871 & 0.764 \\
\geogap{} (ours)    & Full method                     & \textbf{0.935} & \textbf{0.801} \\
\bottomrule
\end{tabular}
\end{table}

\begin{figure}[ht]
\centering
\includegraphics[width=0.75\textwidth]{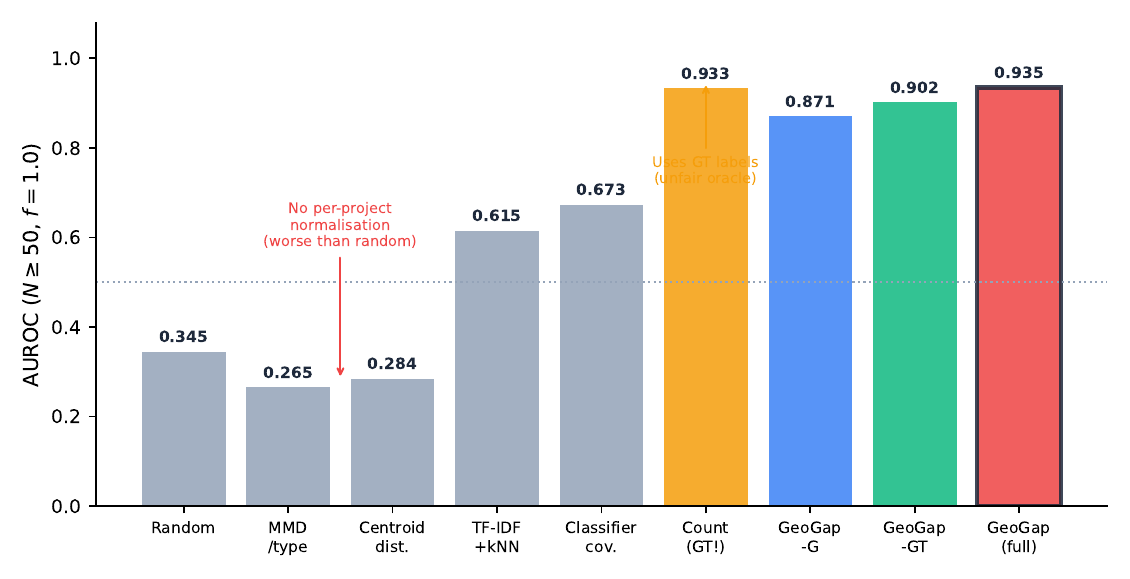}
\caption{Baselines ($N \geq 50$, $f{=}1.0$). Methods without per-project normalisation score below random. \geogap{} matches the GT oracle without annotation.}
\label{fig:baselines}
\end{figure}

\textbf{Validated findings:}
\begin{enumerate}[nosep]
  \item \textbf{Per-project normalisation is essential.} MMD (0.265) and centroid distance (0.284) score \emph{below} random (0.345). Without the per-project baseline $\varphi_0(x)$, raw distances cannot distinguish genuine gaps from inherently sparse regions---the very problem Section~\ref{sec:geo} motivates.
  \item \textbf{Neural embeddings outperform bag-of-words.} \geogapg{} (0.871) vs.\ TF-IDF (0.615): $+42\%$ relative improvement with the same $k$-NN pipeline, demonstrating that contextual embeddings capture semantic structure TF-IDF misses.
  \item \textbf{Geometric scoring beats classify-and-count.} \geogapg{} (0.871) vs.\ classifier coverage (0.673). The classifier approach (inspired by NoRBERT~\citep{hey2020norbert}) predicts types then counts; its 22\% classification error propagates into the coverage estimate.  Operating in continuous embedding space avoids this.
  \item \textbf{\geogap{} matches the ground-truth count oracle.} \geogap{} (0.935) vs.\ GT counting (0.933) for $N \geq 50$---\emph{without requiring human annotation}.  The count baseline uses ground-truth type labels (an unfair advantage unavailable at deployment); the population score $\Psi_{\mathrm{pop}}$ captures the same signal via Gibbs soft assignment.
\end{enumerate}

\subsection{Summary}

\begin{table}[ht]
\centering\small
\caption{Consolidated findings.}
\label{tab:summary}
\begin{tabular}{@{}p{3.8cm}ccl@{}}
\toprule
\textbf{Experiment} & \textbf{AUROC} & \textbf{Verdict} & \textbf{Key finding} \\
\midrule
Type-level, $N{\geq}50$     & 0.935 & Strong & Matches GT oracle \\
Type-level, all              & 0.801 & Good   & Small projects degrade \\
Fraction $f{=}0.75$          & 0.48  & Weak   & Partial gaps hard \\
$k$-sensitivity              & $0.94 \to 0.87$ & Robust & \geogapg{} decays to 0.39 \\
Permutation test             & ---   & 9/15   & All large projects pass \\
Cell-level                   & 0.51  & Random & Grid too sparse \\
Holdout                      & 0.50  & Random & Needs larger corpus \\
\bottomrule
\end{tabular}
\end{table}

\section{Inference}\label{sec:inference}

All corpus artifacts are pre-computed. For a new project: (1)~embed; (2)~assign topics; (3)~compute all three scores; (4)~aggregate; (5)~report. Cost: $<$2~seconds. Minimum $N_* \geq 50$ for reliable detection.

\section{Limitations}\label{sec:limits}

\begin{enumerate}[leftmargin=2em, nosep]
  \item AUROC 0.94 for complete removal but ${\sim}0.50$ for partial gaps ($f \leq 0.75$).
  \item Category assignment is 78.3\% accurate; structured errors create phantom gaps.
  \item 621 points across 96 cells is too sparse for cell-level localisation.
  \item Gaps are corpus-relative; uncovered concerns cannot be flagged.
  \item No practitioner validation of detected gaps.
\end{enumerate}

\section{Future Work}\label{sec:future}

Several directions remain open for strengthening and extending \geogap{}.

\paragraph{Comparison with transformer-based gap detection.}
Our method treats the sentence encoder as a frozen feature extractor and operates entirely in the resulting embedding space. An alternative family of approaches trains transformer architectures end-to-end for gap- or anomaly-related objectives---for example, fine-tuning BERT-based classifiers to predict whether a requirement set is ``complete'' relative to a reference taxonomy, or using sequence-to-set models that directly output missing categories. A systematic comparison between our geometry-based approach and such discriminative transformer baselines would clarify when post-hoc geometric analysis suffices and when end-to-end training is necessary, particularly for the partial-gap regime ($f \leq 0.75$) where \geogap{} currently struggles.

\paragraph{LLM-driven gap analysis.}
Large language models can be prompted to analyse a requirements document and enumerate missing concerns in a single forward pass (e.g., ``\emph{Given these requirements, what quality attributes are underspecified?}''). Such zero-shot or few-shot LLM approaches require no corpus construction and can leverage world knowledge beyond the training data. However, they lack the quantitative, per-type scoring that \geogap{} provides, and their outputs are difficult to calibrate or reproduce. A comparative study measuring precision, recall, and inter-run consistency of LLM-driven gap analysis against \geogap{} would establish the relative strengths of each paradigm---and whether combining them (e.g., using \geogap{} scores as structured context for an LLM's analysis) yields further improvement.

\paragraph{Human expert validation.}
All current experiments evaluate the detector against its own synthetic gap definition. The most important missing piece is a \emph{practitioner study}: present \geogap{} reports to requirements engineers working on real projects, measure whether flagged gaps are judged actionable, and compare against expert-identified gaps that the detector misses. Such a study would move evaluation from ``can the detector recover artificially removed types?'' to ``does the detector help engineers write better specifications?''---which is the ultimate measure of practical value.

\paragraph{Domain-adapted embeddings via fine-tuning.}
\geogap{} currently uses a general-purpose sentence encoder (Qwen3-Embedding-0.6B) without any task-specific adaptation. The embedding space therefore reflects general semantic similarity rather than requirement-coverage structure: two requirements may be semantically close but address different quality attributes, or semantically distant but both relevant to the same gap. Fine-tuning the encoder with a contrastive objective designed for gap detection---for instance, pulling same-type requirements closer and pushing cross-type requirements apart, or training with triplets derived from the synthetic injection protocol itself---could reshape the embedding space to amplify the geometric signal that \geogap{} relies on. This is especially promising for the partial-gap regime, where the current frozen embeddings provide insufficient geometric contrast between ``75\% covered'' and ``fully covered'' types.

\bibliography{references}

\appendix

%======================================================================
\section{Dataset Selection: PROMISE NFR}\label{app:dataset}
%======================================================================

The PROMISE NFR dataset~\citep{clelandhuang2006} is a widely used benchmark for software requirement classification containing $N = 625$ requirement sentences from 15 projects, each annotated with a hierarchical category label.

\subsection{Data Format}

Each record $r_i$ is a tuple $r_i = (\texttt{text}_i, y_i^{(1)}, y_i^{(2)}, \texttt{pid}_i)$, $i = 1, \dots, 625$, where $\texttt{text}_i$ is a natural-language requirement sentence, $y_i^{(1)} \in \{\text{F}, \text{NF}\}$ is the Level-1 label (Functional vs.\ Non-Functional), $y_i^{(2)} \in \mathcal{Y}_2$ is the Level-2 sub-class ($|\mathcal{Y}_2| = 12$), and $\texttt{pid}_i \in \{1,\dots,15\}$ is the source project ID. Every L2 label maps deterministically to an L1 label via a parent function $\pi$: $y_i^{(1)} = \pi(y_i^{(2)})$.

\subsection{Label Taxonomy}

The labels form a two-level tree. Level~1 splits requirements into Functional (F) vs.\ Non-Functional (NF). Level~2 provides 12 sub-classes:

\begin{center}\small
\begin{forest}
  for tree={
    grow'=0, child anchor=west, parent anchor=south,
    anchor=west, calign=first,
    edge path={
      \noexpand\path [draw, \forestoption{edge}]
      (!u.south west) +(7.5pt,0) |- (.child anchor)\forestoption{edge label};
    },
    before typesetting nodes={ if n=1 {insert before={[,phantom]}} {} },
    fit=band, before computing xy={l=18pt},
  }
  [root
    [F\; {\scriptsize(Functional, $n{=}255$)}
      [F\; {\scriptsize--- functional behaviour}]
    ]
    [NF\; {\scriptsize(Non-Functional, $n{=}370$)}
      [A\; {\scriptsize--- Availability ($n{=}21$)}]
      [FT\; {\scriptsize--- Fault Tolerance ($n{=}10$)}]
      [L\; {\scriptsize--- Legal ($n{=}13$)}]
      [LF\; {\scriptsize--- Look \& Feel ($n{=}38$)}]
      [MN\; {\scriptsize--- Maintainability ($n{=}17$)}]
      [O\; {\scriptsize--- Operational ($n{=}62$)}]
      [PE\; {\scriptsize--- Performance ($n{=}54$)}]
      [PO\; {\scriptsize--- Portability ($n{=}1$)}]
      [SC\; {\scriptsize--- Scalability ($n{=}21$)}]
      [SE\; {\scriptsize--- Security ($n{=}66$)}]
      [US\; {\scriptsize--- Usability ($n{=}67$)}]
    ]
  ]
\end{forest}
\end{center}

\subsection{Class Distribution}

The dataset is imbalanced at both levels. The largest class (F, $n{=}255$) is $255\times$ larger than the smallest (PO, $n{=}1$). At L1, the split is NF\,:\,F $=$ 370\,:\,255 $\approx$ 59\,:\,41. Five classes have $n_k < 25$, making per-class metrics noisy. These imbalances motivate our use of Macro-F1 (which weights all classes equally) rather than accuracy as the primary classification metric.

\subsection{How We Use This Dataset}

We use PROMISE NFR in two stages: \textbf{(1)~embedding model selection}---evaluating candidate encoders via zero-shot nearest-prototype classification directly on this dataset; and \textbf{(2)~gap detection experiments}---using the selected model (Qwen3-Embedding-0.6B) and the 12-class taxonomy as the type axis of our coverage analysis. The model selection protocol requires no training data; it tests how well the embedding space separates categories from label semantics alone.

%======================================================================
\section{Embedding Model Selection}\label{app:model-selection}
%======================================================================

\subsection{Motivation}

Our gap detection pipeline operates on sentence embeddings $\mathbf{x}_i = E(r_i) \in \mathbb{R}^d$. The quality of every downstream result depends on the encoder $E$. General-purpose benchmarks (MTEB) rank models across diverse tasks, but a high MTEB score does not guarantee strong performance on short software-requirement sentences with a hierarchical taxonomy. We therefore conducted a \textbf{task-specific model selection} on the PROMISE NFR dataset.

\subsection{Selection Metrics}

\paragraph{Primary: Nearest-Centroid Macro-F1 (zero-shot).}
For each candidate model and description variant $v \in \{\texttt{name\_only}, \texttt{short}, \texttt{detailed}\}$: embed all $N$ requirements, embed the 12 L2 label descriptions as centroids, classify by nearest centroid, report Macro-F1 for the best variant.

\paragraph{Supporting metrics.}
\emph{H-Score}: hierarchy-aware accuracy using normalised tree-path distance.
\emph{ARI/NMI/V-Measure}: $K$-Means clustering alignment with ground-truth labels.
\emph{Linear Probe F1}: supervised logistic regression on frozen embeddings (5-fold CV) as an upper bound.

\subsection{Results}

\begin{table}[ht]
\centering\small
\caption{Model selection on PROMISE NFR. Ranked by NC Macro-F1 (L2).}\label{tab:app-model}
\begin{tabular}{@{}lcccc@{}}
\toprule
Model & NC F1 (L2) & H-Score & ARI & LP F1 \\
\midrule
Qwen3-Embedding-0.6B           & \textbf{.442} & .745          & \textbf{.157} & .715$\pm$.024 \\
Qwen3-Embedding-4B             & .435          & .717          & .088          & .758$\pm$.028 \\
qwen3-embedding:8b             & .427          & \textbf{.790} & .106          & \textbf{.763}$\pm$.040 \\
stella\_en\_400M\_v5            & .402          & .617          & .095          & .757$\pm$.036 \\
gte-large-en-v1.5              & .400          & .711          & .126          & .704$\pm$.061 \\
nomic-embed-text-v1.5          & .389          & .643          & .077          & .729$\pm$.068 \\
e5-large-v2                    & .373          & .660          & .069          & .744$\pm$.038 \\
bge-large-en-v1.5              & .353          & .709          & .141          & .723$\pm$.037 \\
all-MiniLM-L6-v2               & .296          & .578          & .078          & .647$\pm$.044 \\
\bottomrule
\end{tabular}
\end{table}

\paragraph{Key findings.}
(1)~Qwen3-Embedding-0.6B leads on the primary metric (NC F1 = 0.442) and clustering alignment (ARI = 0.157).
(2)~Within the Qwen family, 0.6B $>$ 4B $>$ 8B on NC-F1, while 8B is strongest on LP-F1/H-Score; model size helps some space properties but not all.
(3)~The \texttt{short} description variant is best for 10 of 11 models.
(4)~The supervised--zero-shot gap remains large (0.442 vs.\ 0.763), confirming label-description matching captures only part of the class structure.

\paragraph{Selection.} For downstream experiments requiring zero-shot nearest-centroid scoring, we select \textbf{Qwen3-Embedding-0.6B} ($d = 1024$).

%======================================================================
\section{Preliminary Evaluation of the Embedding Space}\label{app:prelim}
%======================================================================

Before designing the gap detection algorithm, we conducted three experiments to characterise the embedding space and validate that it provides a suitable geometric foundation.

\subsection{Experiment A: Semantic Alignment (Zero-Shot Classification)}

\paragraph{Question.} Does the embedding space organise requirements in a way that aligns with human-defined categories?

\paragraph{Protocol.} \emph{Top-down (Test a):} Nearest-centroid classification using label description embeddings as centroids. Tested at both L1 (F/NF) and L2 (12 sub-classes) with three description variants.
\emph{Bottom-up (Test b):} $K$-Means, HDBSCAN, and Agglomerative Hierarchical Clustering on raw embeddings, compared to ground-truth labels via ARI, NMI, V-Measure, and Purity.

\begin{table}[ht]
\centering\small
\caption{Semantic alignment: nearest-centroid classification.}\label{tab:app-align}
\begin{tabular}{@{}lccccc@{}}
\toprule
Level & Best variant & Accuracy [95\% CI] & Macro-F1 & $\kappa$ & H-Score \\
\midrule
L1 & detailed & 0.502 [0.464, 0.542] & 0.481 & 0.102 & --- \\
L2 & short    & 0.467 [0.429, 0.507] & 0.442 & 0.393 & 0.745 \\
\bottomrule
\end{tabular}
\end{table}

\begin{table}[ht]
\centering\small
\caption{Clustering alignment at L2 ($K_2 = 12$). $p$-values from permutation test ($n = 1000$), BH-corrected.}\label{tab:app-cluster}
\begin{tabular}{@{}lccccl@{}}
\toprule
Algorithm & ARI & NMI & V-Measure & Purity & $p$(ARI) \\
\midrule
$K$-Means          & \textbf{0.157} & \textbf{0.323} & \textbf{0.323} & 0.594 & $<$0.001 \\
AHC (Ward)         & 0.119          & 0.292          & 0.292          & 0.558 & $<$0.001 \\
HDBSCAN            & $-$0.041       & 0.038          & 0.038          & 0.408 & 1.000 \\
TF-IDF + $K$-Means & $-$0.011       & 0.193          & 0.193          & 0.454 & --- \\
Random             &  0.001         & 0.053          & 0.053          & 0.408 & --- \\
\bottomrule
\end{tabular}
\end{table}

\paragraph{Findings.}
(1)~L2 accuracy (0.467) is nearly $6\times$ the 12-class chance baseline (0.083), with $\kappa = 0.393$ (``fair'' agreement). H-Score = 0.745 confirms most errors are within the same taxonomy branch.
(2)~$K$-Means at L2 achieves ARI = 0.157 ($p < 0.001$), substantially above random, confirming the embedding space has natural clusters that align with the label taxonomy.
(3)~Embedding-based clustering outperforms TF-IDF by 67\% (NMI: 0.323 vs.\ 0.193).
(4)~HDBSCAN fails (ARI = $-0.041$), consistent with concentration of measure in $d = 1024$.

\paragraph{Motivation for \geogap{}.} These results show the embedding space encodes meaningful requirement-type structure, but imperfectly (78.3\% centroid accuracy, ARI = 0.157). A naive classifier applied to this space would propagate the 22\% error rate. This motivates the geometric $k$-NN approach: instead of classifying and counting, \geogap{} operates directly in the continuous embedding space, avoiding hard classification errors entirely.

\subsection{Experiment B: Coverage Analysis}

\paragraph{Question.} Can the embedding space quantify how well different requirement sources cover each other?

\paragraph{Protocol.} Embed three text sets---raw documentation, ground-truth (GT) requirements, and LLM-extracted requirements---and compare their distributions using Wasserstein-1, Sliced Wasserstein, MMD$^2$, and point-level coverage metrics (ANND, AUCC).

\paragraph{Findings (synthetic data, $N \leq 16$).}
(1)~LLM outputs are distributionally closest to GT (lowest $W_1$, $\text{MMD}^2$, ANND).
(2)~46.7\% of GT requirements are ``hidden'' (far from any raw documentation chunk); the LLM recovers 85.7\% of these.
(3)~The embedding geometry provides a natural framework for measuring coverage---the per-point nearest-neighbour distances that \geogap{} uses for gap detection originate from this observation.

\subsection{Experiment C: Trajectory Analysis}

\paragraph{Question.} Does the embedding model's classification converge as more of a requirement's text is revealed?

\paragraph{Protocol.} Inspired by the Landscape of Thoughts framework, embed progressively longer prefixes of each requirement and track the Gibbs-distribution probability vector over 12 category centroids at each step.

\paragraph{Findings ($n = 100$ requirements).}
(1)~Mean uncertainty is 2.483, within 0.08\% of maximum entropy ($\log 12 = 2.485$), indicating the Gibbs distribution is essentially uniform at $T = 1.0$.
(2)~Correct trajectories converge $6\times$ faster (speed = 0.166 vs.\ 1.000) and are 53\% more consistent (0.874 vs.\ 0.571) than incorrect ones.
(3)~The near-uniform uncertainty motivates our calibrated temperature $T^* = 0.021$ for the population count score $\Psi_{\mathrm{pop}}$: at $T = 1.0$, Gibbs assignment is too flat to be useful; at $T^* = 0.021$, it matches the 78.3\% hard-assignment accuracy.

\subsection{Cross-Experiment Synthesis}

These preliminary experiments establish three facts that motivate \geogap{}:
\begin{enumerate}[nosep]
  \item The embedding space encodes requirement-type structure (L2 Macro-F1 = 0.442, ARI = 0.157), but imperfectly---motivating geometric coverage analysis over hard classification.
  \item Nearest-neighbour distances in this space naturally capture coverage relationships---motivating the $k$-NN formulation.
  \item Gibbs soft assignment at calibrated temperature provides annotation-free type counting---motivating the population score $\Psi_{\mathrm{pop}}$.
\end{enumerate}

\end{document}